\documentclass[10pt]{article}

\usepackage{amsmath}
\usepackage{amssymb}

\usepackage{graphicx}

\usepackage{cite}

\usepackage{color} 

\usepackage[labelfont=bf,labelsep=period,justification=raggedright]{caption}

\makeatletter
\renewcommand{\@biblabel}[1]{\quad#1.}
\makeatother

\date{}

\begin{document}

\begin{flushleft}
{\Large
\textbf{Determinants of the Pace of Global Innovation in Energy Technologies}
}
\\
Luis M. A. Bettencourt$^{1,2,\ast}$, 
Jessika E. Trancik$^{1,3,\ast, \ast\ast}$, 
Jasleen Kaur$^{4}$
\\
\bf{1} Santa Fe Institute, 1399 Hyde Park Road, Santa Fe, NM 87501, USA
\\
\bf{2} Theoretical Division T5, MS B284, Los Alamos National Laboratory, Los Alamos, NM 87545, USA
\\
\bf{3} Engineering Systems Division, Massachusetts Institute of Technology, Cambridge, MA 02139, USA
\\
\bf{4} Center for Complex Networks and Systems Research, School of Informatics and Computing, Indiana University, Bloomington, IN 47406, USA
\\
$\ast$ These authors contribute equally.
\\
$\ast\ast$ Corresponding author:  trancik@mit.edu
\end{flushleft}

\section*{Abstract}
Understanding the factors driving innovation in energy technologies is of critical importance to mitigating climate change and addressing other energy-related global challenges. Low levels of innovation, measured in terms of energy patent filings, were noted in the 1980s and 90s as an issue of concern and were attributed to low investment in public and private research and development (R\&D). Here we build a comprehensive global database of energy patents covering the period 1970-2009 which is unique in its temporal and geographical scope. Analysis of the data reveals a recent, marked departure from historical trends. A sharp increase in rates of patenting has occurred over the last decade, particularly in renewable technologies, despite continued low levels of R\&D funding. To solve the puzzle of fast innovation despite modest R\&D increases we develop a model that explains the nonlinear response observed in the empirical data of technological innovation to various types of investment. The model reveals a regular relationship between patents, R\&D funding, and growing markets across technologies, and accurately predicts patenting rates at different stages of technological maturity and market development. We show quantitatively how growing markets have formed a vital complement to public R\&D in driving innovative activity; these two forms of investment have each leveraged the effect of the other in driving patenting trends over long periods of time.

\section*{Introduction}
Over the last century the global energy sector has seen several bursts of technological innovation \cite{Margolis99, NAS09, Nemet07b, Margolis99b, Newell11}. The energy crises of the 1970s generated much enthusiasm for renewable and other alternative energy technologies and suggested an imminent technological transition. However, as oil prices fell in the mid-1980s, alternative energy technologies were no longer favored and several decades of low research investment ensued. Coincident with these trends, rates of patenting stagnated during 1980-2000  \cite{Margolis99, Nemet07b, Margolis99b}. The observed correlation between total (public and private) R\&D and patenting in the US over the period of 1970-2003 suggested that this slowdown in innovation was the direct result of disinvestment in research \cite{Margolis99, Nemet07b, Margolis99b}.

More recently, due to climate change and energy security concerns \cite{Caldeira03, team07, Fri03, Hoffert98}, interest in alternative energy technologies has been growing once again \cite{UN10, Pew10, Dechezlepretre11}. To assess the current state of the field we built a comprehensive database of energy patents filed across the world (roughly 73,000 patents covering 1970-2009, see Table S1). We aggregated patents geographically and by technology, building on the methods of references \cite{Margolis99, Nemet07b, UN10, Margolis99b} but expanding their temporal and geographical scope. We also assembled data on production levels and public research and development funding for nations and technologies over time. 

We analyzed temporal trends in patents and funding in order to understand the level of innovative activity across technologies and nations and to probe the drivers of these trends. Patents provide an unparalleled measure of the location and intensity of innovative activity\footnote{The innovative activity signaled by a patent typically occurs at the transition point between invention and innovation \cite{Ruttan59}.}\cite{Jaffe02, Johnstone10, Strumsky10, Nuvolari11}. When coupled with appropriate verification that changes in patent counts over time are not artifacts of changes to policies regulating intellectual property or changes to patent quality (see Figures S1 and S2), a comprehensive patent database is a powerful tool for investigating the determinants of innovative activity. Our approach involves studying both national and global trends for various energy technologies, and searching for a simple and general explanatory model in order to avoid overfitting the data and to maximize predictive power. We develop a model that explains the observed trends in energy patenting by capturing the global nature of innovation and the persistence of knowledge over long periods of time. 


\section{Results}

\subsection{Temporal and Regional Trends in Patents and Funding}

The empirical evidence points to a pronounced increase in patenting in energy technologies over the last decade (Figure 1), despite traditional investment---private and public R\&D---not rising commensurately.  (Public R\&D data at the global level is more readily available than private R\&D \cite{Gallagher11}. Surveyed private R\&D remained low in years for which data are available. See reference~\cite{Nemet07b} and the Supplementary Information.) 

A more detailed examination of the structure of these trends sheds additional light on the phenomenon. First, we observe marked differences in patenting rates across technologies. Although almost every technology sector has experienced increases in patenting over the last decade, there are notable differences which hold across nations. Renewable energy technologies---especially solar and wind---are growing most rapidly while patenting in nuclear fission has remained low despite sustained high levels of public investment (Figure 1). 

\begin{figure*}
\vspace{2em}
\centering
\hspace{-2em}
\includegraphics[width=0.6\textwidth]{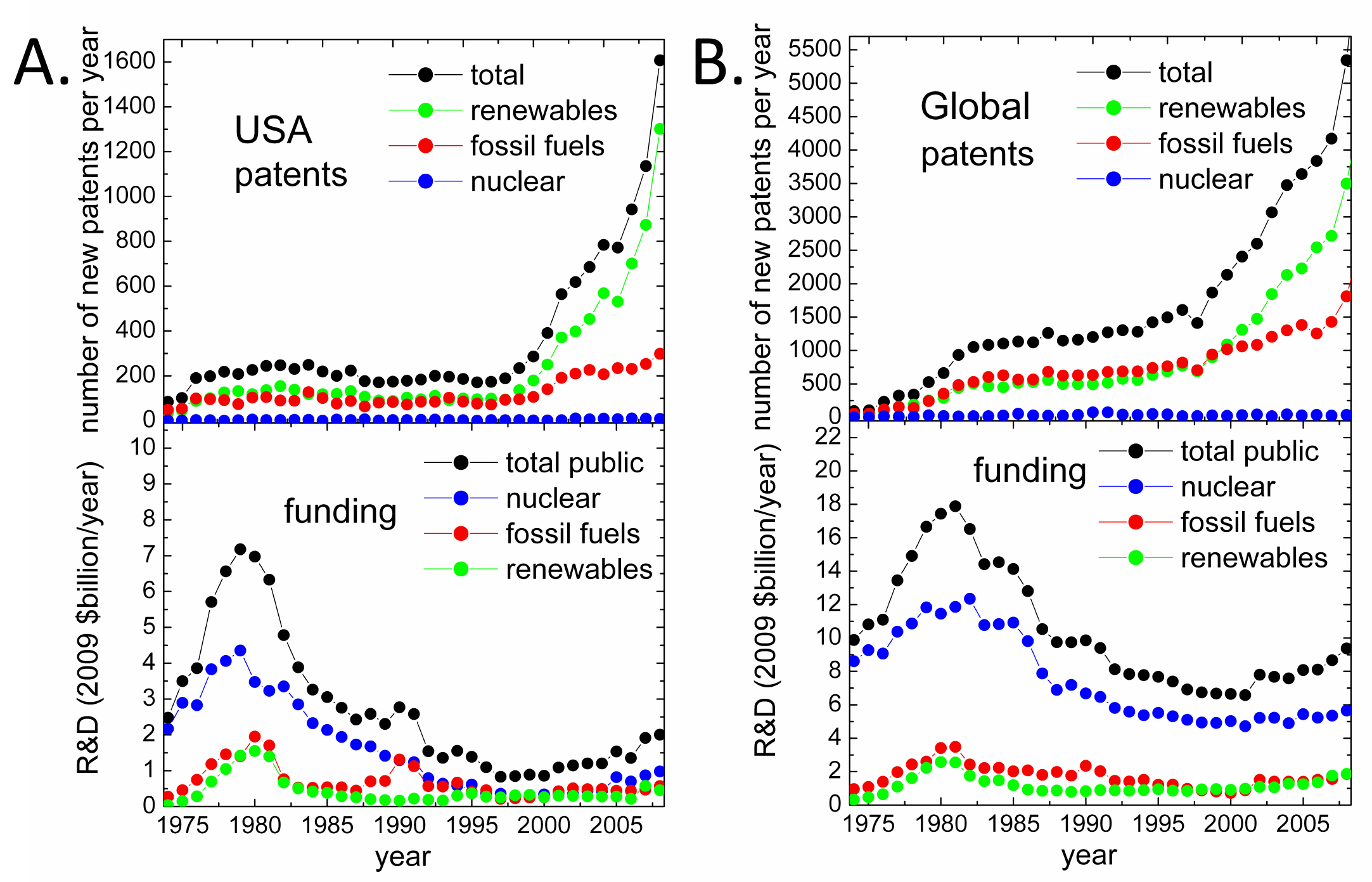} 
\caption{Temporal trends in energy patents: 
A. Time series for US energy patents over the period 1970-2009 show rapid growth over the last decade (upper panel), with the greatest increases in renewables patents. Public funding (lower panel) for energy R\&D increased dramatically in the 70s and 80s but then remained relatively constant. B. A similar trend in patenting is seen when aggregating globally (upper panel). Funding for public R\&D among IEA countries (lower panel) shows only modest increases in recent years.}
\label{trends_patents}
\end{figure*}

Solar and wind patents are growing the fastest among the technologies studied, with average annual growth rates during 2004-2009 of 13\% and 19\%, respectively, while the corresponding number for all energy technologies was 11.9\%. According to the World International Patent Office \cite{WIPO10} the worldwide figure for energy was 4.6\% between 2001-2005, which has come to match or surpass growth rates for Òhigh-techÓ sectors such as semiconductors (4.9\%) and digital communication (3.0\%). The fraction of all patents accounted for by energy, and by those in solar and wind technologies in particular, has increased significantly in recent years (Figure S1). Moreover, patent citations (a proxy for patent impact or quality \cite{Jaffe02,Hall01}) have remained steady or grown slightly. Thus, there is no quantitatively observable indication for patent inflation (Figure S2), though more time is needed to accumulate citations for the newest patents \cite{Jaffe02,Hall01}.  

Second, we find increasing rates of patenting across the world but also differences in regional priorities (Figure 2). Although the numbers of patents filed in intellectual property offices around the world vary because of different standards and scope, the distribution of these patents across technologies allow us to probe technological preferences in each region (Figure 2). The Japanese patent office is the leader in terms of total (cumulative) patents filed for all energy technologies apart from coal, hydroelectric, biofuels, and natural gas (Figure 2). The European Patent Office has shown a downturn in fossil fuel patent applications over the last decade (Figure S3), particularly in coal. China is now logging more energy patents per year than the European Patent Office and growing much faster than any other nation (Figure S4). More patents related to coal technologies have been filed in China than any other nation, and China now comes a close second to Japan in terms of cumulative wind patents (Figure 2). 

These trends contradict a picture of patenting in energy technologies that is primarily driven by inputs in public R\&D investment \cite{Margolis99,Nemet07b,Margolis99b}. They  point to the relevance of opportunities resulting from the growth of markets, which drives an increase in both explicit private R\&D funding and other forms of investment that generate innovative activity.  Consequently, a new framework is needed to account simultaneously for the old (pre-2000) and new (post-2000) patterns of innovation across different energy technologies \cite{Henderson11,Pew10,WIPO10,Gallagher06,Gompers01, Hall02, Kortum00, Newell10}.

\begin{figure*}
\vspace{2em}
\centering
\hspace{-2em}
\includegraphics[width=.9\textwidth]{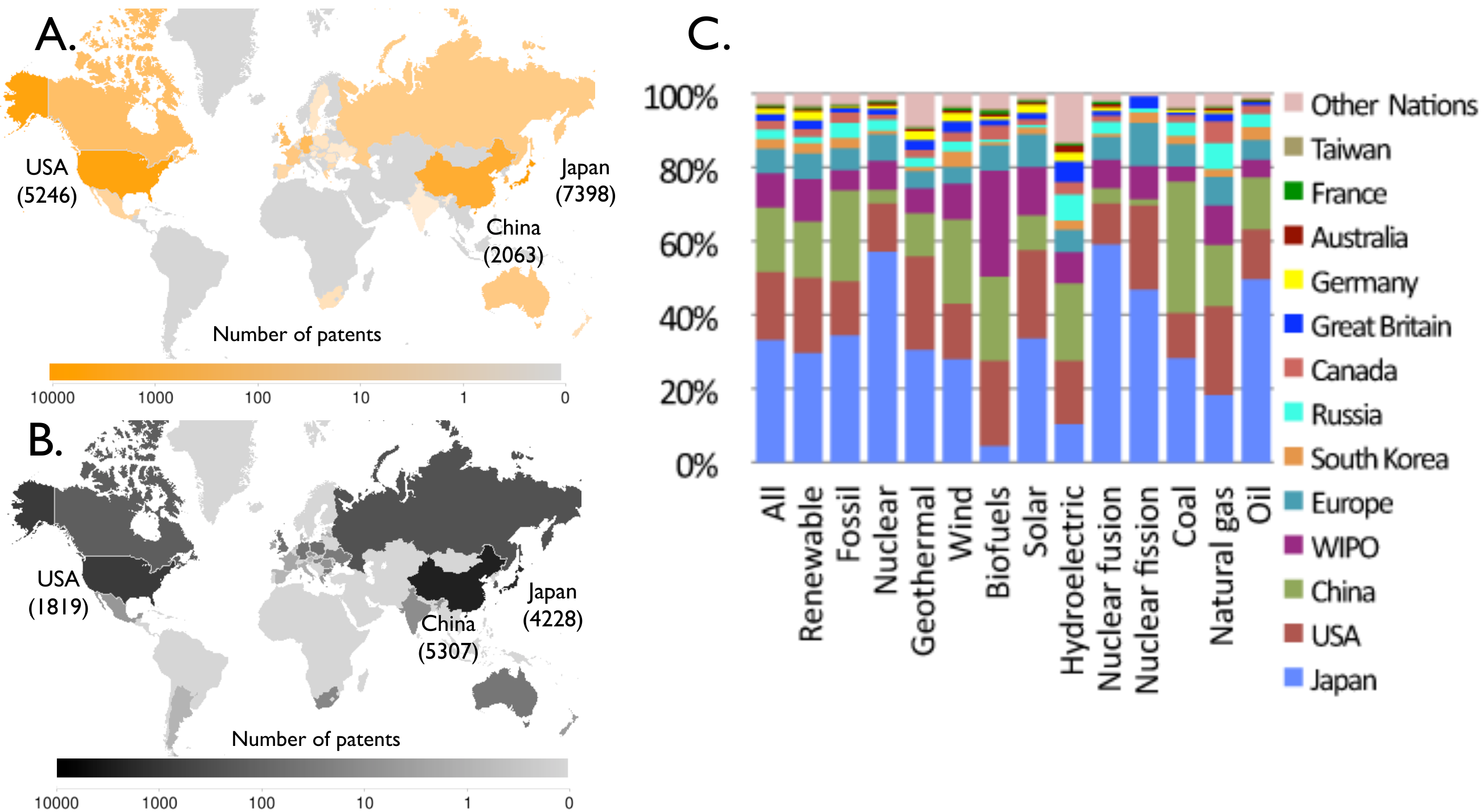} 
\caption{Regional patterns in energy patents. A. World map of cumulative patents in photovoltaics (solar). Japan is the leading nation in terms of patent numbers, followed by the US and China (patent numbers through 2009 are shown in brackets). In addition there are 1951 patent applications in the European Union (EPO) and 2882 worldwide (not shown). (Patent applications in wind technologies are distributed similarly and therefore not shown. Japan is again the leading nation, followed by China and the US. There are also 702 patent applications in the European Union and 1484 worldwide.) B. World map of patent applications in coal-based electricity. China is the leading nation, followed by Japan, and the US. There are 887 patents filed in the European Union and 620 worldwide. C. Cumulative patents submitted by technology and nation indicate the dominance in terms of patent counts of certain countries, and also differences across countries in the relative importance placed on each technology.}
\label{regional_patterns}
\end{figure*}

\subsection{Model of Patent Growth and Technological Innovation}

We begin with a conceptual scheme by hypothesizing that many technologies experience similar phases of development \cite{Henderson11}, Figure 3. First, new technologies require a developmental period during which markets are small or inexistent. At this stage basic R\&D investments are essential, usually through public investments or niche ÒmarketsÓ such as military applications (Figure 3A). There are many examples of important technologies that went through this stage of development from integrated circuits, computers and the internet, to cell phones and nuclear energy \cite{Henderson11,Trancik06}. This stage can be slow but as long as there is some investment the cumulative and multiplicative effects of knowledge creation will tend to generate new solutions \cite{Henderson11,Koh08}. 

For successful technologies markets then materialize and grow. At this second stage of development, Figure 3B, new and established companies acquire the means and motivation to invest in improving the product's performance, price, and reliability, if the growth rates of these products are high enough to exceed the market's average \cite{Damodaran11}.  A virtuous self-reinforcing cycle connecting product improvements to expanding markets is thus created that sustains technological improvement, and the technology takes off. Recent improvements in computer and mobile device technologies are in this category. 

Translating these general ideas into a mathematical model allows us to account for the empirical trends described above. We draw on a tradition of work in technological innovation and patenting, mostly in economics \cite{Acs02, Jaffe86}, and specifically on the well-known ideas of Griliches \cite{Griliches90}. Specifically, we draw on the qualitative insights of this early work but introduce several important modifications to develop a mathematical model that is consistent with empirical evidence, as described further below. This leads to a new model that is consistent with data on the development of energy technologies. 

The fundamental quantity in our model and in previous work is knowledge, $K$; this is difficult to quantify because $K$ is generally an unobservable quantity. However, its observable inputs are quantifiable in the form of public R\&D, $R$, and private investment that grows with production levels, $C$. ($C$ is a proxy for private investments in R\&D and other private efforts that lead to knowledge creation. Note here that for the years and locations for which private R\&D is available, such as in the US up to 2003, the growth in documented private R\&D does not explain the large growth in patents. $C$, together with public R\&D, is a stronger determinant of patenting levels than total public and private R\&D.) The measurable outputs of $K$ are patents, $P$, and increases in technology performance, which affect production levels, $C$. We can then short-circuit $K$ in terms of its inputs and outputs (Figure 3). GrilichesÕ original model \cite{Griliches90} assumed a linear relationship among these quantities, but empirical observations falsify this choice. To account for observed trends we introduce a nonlinear function relating cumulative quantities 
\begin{align}
P = P_{0}R^{\alpha}C^{\beta}, 
\end{align}
where $P_{0}$ is a constant, measuring the productivity in excess of that due to changes in the inputs $R$ or $C$. We provide a discussion of this relationship in the SI, emphasizing its dynamical origin and discussing additional factors involved. 

Relations such as Eq. (1) are common in socioeconomic phenomena where it is often the proportional rates of change of outputs to inputs that are related linearly, not the quantities themselves. Examples include aggregate production functions \cite{Barro04} and two-factor technology learning curves \cite{Jamasb07}. The production of new patents, $dP/dt$, is written in terms of new R\&D investments $dR/dt$ and expanding markets $dC/dt$ as
\begin{align}
\frac{1}{P}\frac{dP}{dt} =\frac{\alpha}{R}\frac{dR}{dt} + \frac{\beta}{C}\frac{dC}{dt}, 
\end{align}
where the proportionality factors $\alpha$ and $\beta$ are dimensionless numbers (exponents) known in economics as elasticities. Larger elasticities correspond to greater patent output in terms of each additional unit of input in $R$ or $C$. Because of the cumulative nature of $C$ and $R$, even if either input ceases to grow its cumulative base will continue to enhance the value of new investment in the other. Past public R\&D funding continues to boost the value of new private investments and vice-versa.  Note that here we are relying on a conceptual model (Figure 3) in addition to the mathematical relationship shown above. Together they propose a causal relationship, which captures the endogenous feedbacks between variables.

Figure 4 shows the fit of model Eq.~(1) to global data for the two largest technologies in terms of patenting: solar and wind. We focus on global data because of the international nature of innovation in these technologies. It is common for a technology to be, for example, developed by a US firm, patented and manufactured in China, and sold and installed in Europe. We also show the fit to US coal data, where innovation is expected to be more nationally centered.  We observe that the model accounts for the trends in patenting ($R^2 >0.98$ and very small $p$-values, Tables S2-7) across the three technologies and the entire period of 1974-2009. As expected, a fit based exclusively on public R\&D investment (open squares) fails to describe the data and in particular results in the wrong curvature, underestimating the recent rise in patenting rates in all three technologies. Thus, accounting for both public R\&D and market growth is essential for predicting rates of innovation. The absence of a time dependence of $P_0$ shows that the effect of R\&D investments and market growth on technological innovation in each energy technology has remained essentially unchanged over decades, and that changes in rates of patenting are the result of evolving levels of these inputs. (In the Supplementary Information, we also present the fit of Eq.~(2) to global data to further verify these results.)

Most technologies show greater sensitivity to market growth ($\beta > \alpha$) than to public R\&D investments, though for wind the two contributions are similar with $\alpha \gtrsim \beta$. Wind technologies are also peculiar in another way as patenting trends are best predicted by taking production, $C$, with a negative time lag, that is 3-4 years in the future. (This may reflect particularities about the advanced planning of large wind installations \cite{Pew10}, though further, detailed analysis is needed.) For all three technologies, the model best fits the data when the effects of public R\&D are seen in the same year or one year before the date of patent application (see Figure 4 and Table S2-S4). Other investments, such as venture capital \cite{Pew10, WIPO10, Gompers01, Hall02, Kortum00}, not accounted for by $R$ and $C$ may also play a role (see derivation and discussion in SI) but their rates of change must be highly correlated to the former in order not to produce diverging temporal trends in $P$. 

\begin{figure}
\vspace{2em}
\centering
\hspace{-2em}
\includegraphics[width=.4\textwidth]{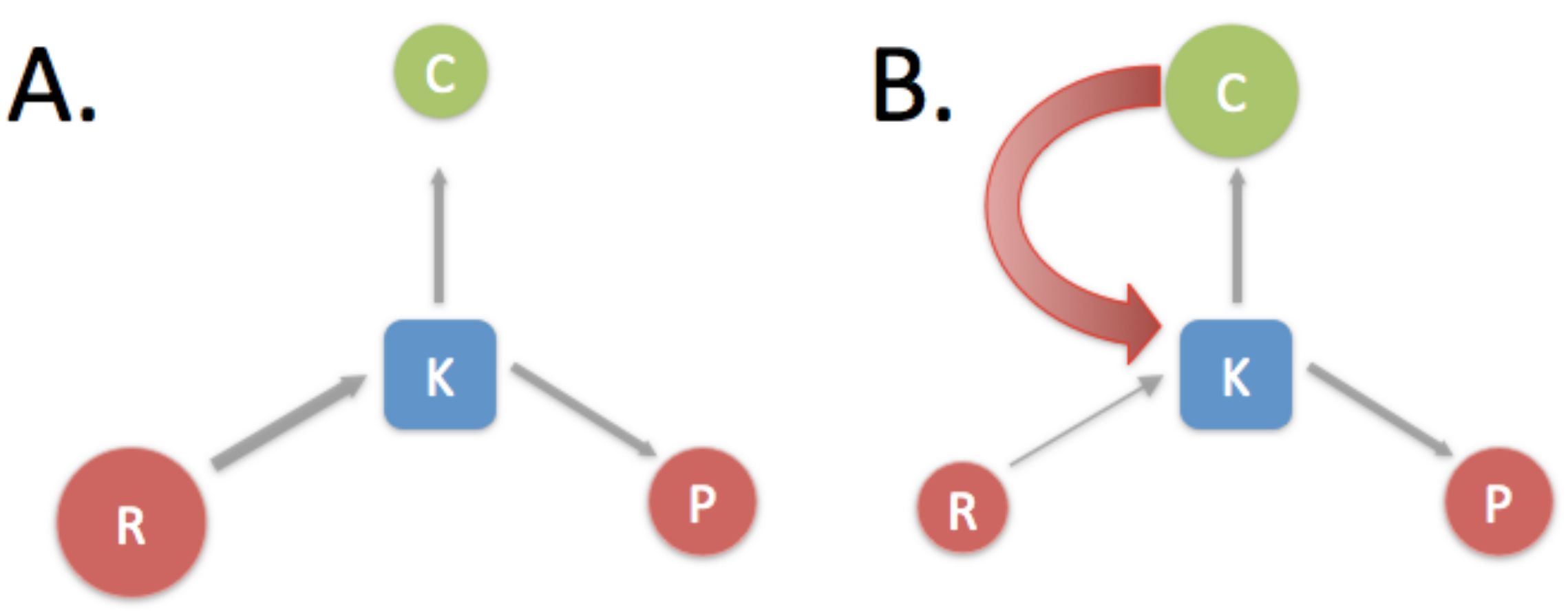} 
\caption{Schematic model of technology evolution. 
Quantities in circles ($C$, $P$, $R$) are observable inputs and outputs of innovation processes, while knowledge $K$ is not observable. A. At early stages of a technology, markets, $C$, are typically very small and public R\&D investments $R$ are essential for generating new knowledge $K$ and resulting patents $P$ and product improvements. This has been the case for solar, wind and other energy technologies in the last few decades. B. As markets materialize, investments in innovation are increasingly driven by market growth (curved arrow), until this mechanism takes over. Market growth may be driven by public policy. At some point the technology may enter a cycle of rapid innovation (it takes off) and public R\&D investment is no longer a dominant driver. Public R\&D investments in innovation and those driven by market expansion have effects that are multiplicative, with each providing a base multiplier for the other. Any public R\&D investment is highly leveraged by market driven investments as technologies develop towards stage B, as is presently occurring with several energy technologies such as solar and wind.
}
\label{schematic}
\end{figure}

Thus, the model demonstrates that a virtuous innovation cycle formed by R\&D support and market growth can account for the sharp increase in energy patenting observed in recent years. (In the context of energy, both have been heavily dependent on public policy \cite{Popp10}.) This explanation holds across diverse technologies, despite large differences in maturity and market size. The model also holds across different eras---the energy crises and attendant increase in research funding, the stagnant decades, and recent years in which concerns about climate change have intensified. The model applies to wind, solar and coal but also to other technologies such as nuclear fission, which shows low patenting rates concurrent with slow market growth. 

\section{Discussion}

We have shown, based on an extensive survey of patents filed throughout the world, that the pace of technological innovation in energy technologies has shown large variations over time. In recent years there as been a boom in energy innovation, as measured by patents, which is dominated by solar and wind conversion technologies. These trends over time and across technologies cannot be explained by public R\&D funding alone but, as we have shown here, can be accounted for by the combined effects of public investments in R\&D and a fast rate of growth in markets for these technologies.  

Our quantitative analysis was carried out primarily at a global level because commercial transactions and knowledge transfers across nations are important and common in energy technologies~\cite{Dechezlepretre11}. These transfers of knowledge are evident in the character of product development (visible in patent assignees), manufacturing, and installation, which commonly involve a number of different firms and nations.  Because of the international nature of this process, analysis of narrower (e.g. firm or nation specific) datasets may fail to capture the innovation and commercial cycle. Our quantitative results support this global model of innovation. However, we also highlight underlying national trends (e.g. the high growth rates in Chinese patents) as these shed light on the degree to which nations are able to capitalize and benefit from innovation in energy. 

We find that both market-driven investment and publicly-funded R\&D act as base multipliers for each other in driving technological development at the global level. We also find that the effects of these investments persist over long periods of time, supporting the notion that technology-relevant knowledge is preserved. 

Our results suggest that how markets are created, which varies significantly temporally and geographically~\cite{Johnstone10, Gompers01, Nemet10}, is not a key determinant of patenting levels at a global scale. To the extent that markets for these technologies grow fast enough, economic opportunity drives an increase in patenting and knowledge creation. It is important to emphasize that the growth of markets for low-carbon energy technologies, which improve on an aspect of performance (carbon emissions) not commonly captured by market price, has depended strongly on public policy. We also note that policies are likely needed to fund research and incentivize market growth further until these technologies become cost-competitive and can take off on their own. 

This suggestion of the dependence of global patenting trends on the aggregate scale of research funding and markets, rather than the details of policy instruments and other incentives, is important because of the diversity of public policies at finer geographical scales and limited ability to coordinate these policies across national borders. Similarly, the apparent persistence of knowledge over long time periods is an important result given the variability (and lack of continuity) in policies over time.  

We focus on broad classes of technologies which endure over time in order to capture long-term patterns of technological innovation and their principal drivers. These results do not, however, undermine the importance of carefully tailoring policies to support the most important innovations (e.g. development of photovoltaics based on earth-abundant materials). For technologies which are further away from cost-competitiveness the detailed nature of innovation---fundamental or applied---will be especially critical for their long-term prospects and certain policy instruments should target these particular aspects. Further detailed studies of the effects of policy instruments on the nature of innovation are needed to complement our analysis and address these issues. 

The data and model developed here focus on the immense and complex energy sector. The general framework developed may, however, prove relevant to the development of other technologies and to general theories of technological innovation.

\begin{figure}
\vspace{2em}
\centering
\hspace{-2em}
\includegraphics[width=0.4\textwidth]{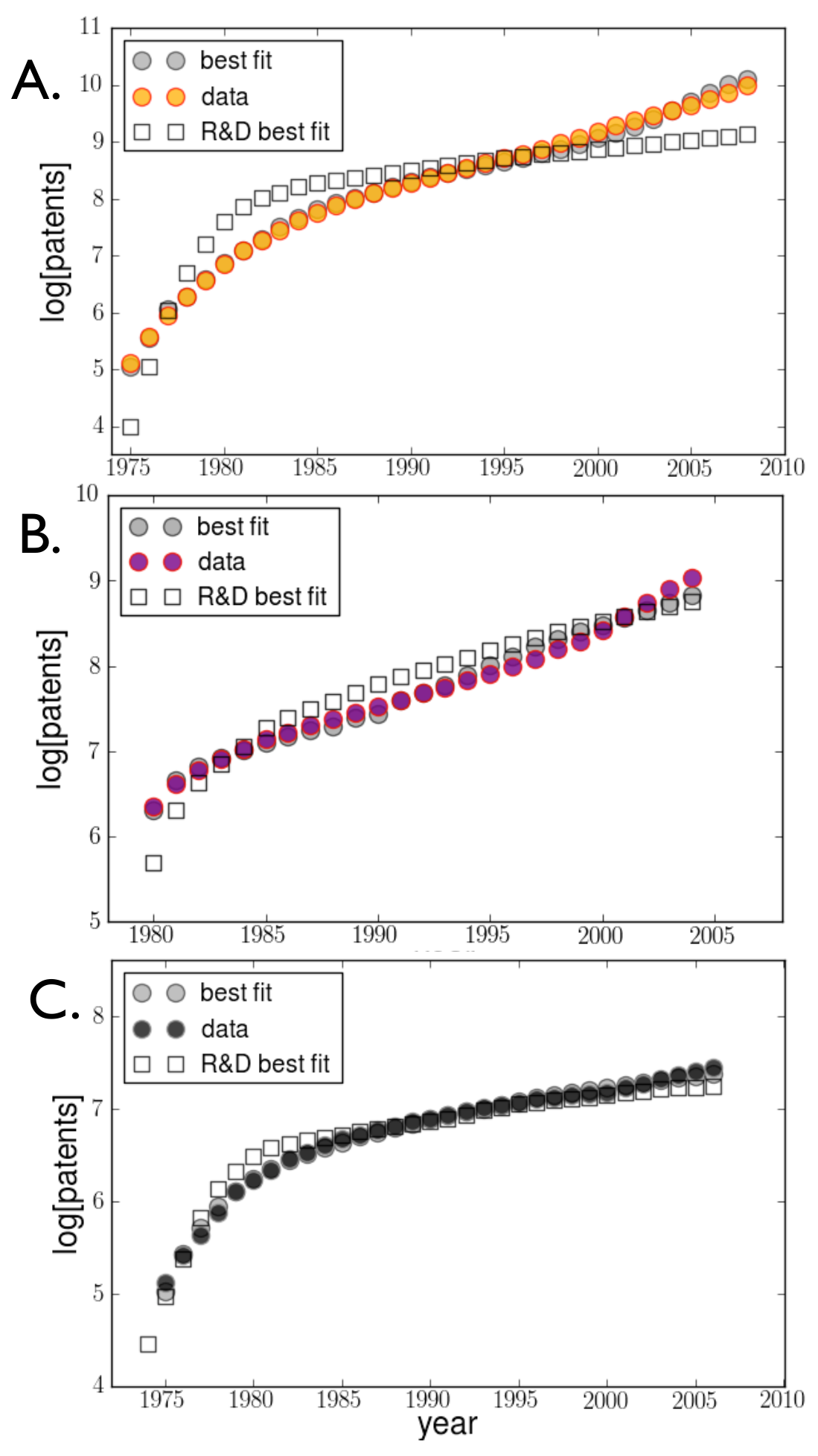} 
\caption{Patenting growth, R\&D investment and market expansion for A. Solar (photovoltaic) worldwide, B. Wind worldwide and C. Coal, US only. Innovation, proxied by (cumulative) patent applications, is well described (grey dots) by a production function that includes both the effects of cumulative production and cumulative public R\&D investment for all technologies. This relationship holds at various stages of maturity. A restricted Eq. (1) accounting for public R\&D investments alone gives a poor description of the trends observed (open squares), failing in particular to account for the uptick in patenting (and upwards curvature) in wind and solar technologies over the last decade. Full details of the fit parameters and statistical goodness of fit are given in the SI. The best fit for solar has $\alpha$=0.22, $\beta$=0.41, $R^2=0.997$ and p-values $\textless$ $e^{-9}$. The best fit for wind has $\alpha$ =0.51, $\beta$ =0.30, $R^2=0.983$ and corresponding p-values$=7.9 e^{-5}$ and $1.4 e^{-9}$, respectively. The best fit for coal has $\alpha$=0.18, $\beta$=0.43, $R^2=0.997$ and p-values $\textless$ $e^{-12}$.  }
\label{fits}
\end{figure}

\section*{Acknowledgments}
This work was partially supported by the National Science Foundation under grant SBE-0738187 (to JET), by the Army Research Office under grant W911NF-12-1-0097 (to LB), and by the Los Alamos National Laboratory LDRD program under grant SGER-0742161 (to LB). We thank Hamed Ghoddusi for comments on this manuscript.

\bibliographystyle{plos}
\bibliography{patents}



\end{document}